\documentclass[12pt]{article}
\usepackage{graphics}
\usepackage{graphicx}

\input{epsf}
\usepackage{cite}
\def\@fmsl@sh#1#2#3{\m@th\ooalign{$\hfil#1\mkern#2/\hfil$\crcr$#1#3$}}
 \def\eq#1\en{\begin{equation}#1\end{equation}}
\def\s[#1,#2]{[#1\stackrel{\star}{,}#2]}
\def\sx[#1,#2]{[#1\stackrel{\star_{x}}{,}#2]}

\newcommand{\nc}{\newcommand}
\nc{\beq}{\begin{equation}}
\nc{\eeq}{\end{equation}}
\nc{\beqa}{\begin{eqnarray}}
\nc{\eeqa}{\end{eqnarray}}

\def\gsim{\mathrel{\rlap{\lower4pt\hbox{\hskip1pt$\sim$}}
    \raise1pt\hbox{$>$}}}       

\begin{document}
\makeatletter
\def\fmslash{\@ifnextchar[{\fmsl@sh}{\fmsl@sh[0mu]}}
\def\fmsl@sh[#1]#2{%
  \mathchoice
    {\@fmsl@sh\displaystyle{#1}{#2}}%
    {\@fmsl@sh\textstyle{#1}{#2}}%
    {\@fmsl@sh\scriptstyle{#1}{#2}}%
    {\@fmsl@sh\scriptscriptstyle{#1}{#2}}}
\def\@fmsl@sh#1#2#3{\m@th\ooalign{$\hfil#1\mkern#2/\hfil$\crcr$#1#3$}}
\makeatother


\title{\large{\bf Colorful quantum black holes at the LHC}}

\author{Xavier Calmet$^{a}$\thanks{Charg\'e de recherches du F.R.S.-FNRS.}
\thanks{xavier.calmet@uclouvain.be}, \
Wei Gong$^{b}$\thanks{wgong@uoregon.edu} \ and \
Stephen~D.~H.~Hsu$^{b}$\thanks{hsu@uoregon.edu} \bigskip
\\
 $^{a}$Catholic University of Louvain,\\
Center for Particle Physics and Phenomenology,\\
2, Chemin du Cyclotron,\\
B-1348 Louvain-la-Neuve, Belgium\\
$^{b}$Institute of Theoretical Science, University of Oregon,\\
Eugene, OR 97403 USA}

\date{June, 2008}

\maketitle

\begin{abstract}
We examine the LHC phenomenology of quantum black holes in models of TeV gravity. By quantum black holes we mean black holes of the smallest masses and entropies, far from the semiclassical regime. These black holes are formed and decay over short distances, and typically carry SU(3) color charges inherited from their parton progenitors. Based on a few minimal assumptions, such as gauge invariance, we identify interesting signatures for quantum black hole decay such as 2 jets, jet + hard photon, jet + missing energy and jet + charged lepton, which should be readily visible above background. The detailed phenomenology depends heavily on whether one requires a Lorentz invariant, low-energy effective field theory description of black hole processes.
\end{abstract}


\newpage

\bigskip

Gravity might be much stronger in the TeV regime than naively expected. Large extra-dimensions \cite{ArkaniHamed:1998rs,Randall:1999ee} or a large hidden sector \cite{Calmet:2008tn} can lead to dramatic modifications of the strength of gravity, potentially reducing the scale of strong quantum gravity to of order TeV, and ameliorating the hierarchy problem. If this is the case, the most striking feature of these models is the prediction that colliders such as the LHC may be able to create small black holes. However, due to improvements in our understanding of the formation of semiclassical black holes in a collider setting \cite{Eardley:2002re}, it now seems very unlikely that semiclassical black holes, e.g., with semiclassical spacetimes,  which decay to many final state particles, will be produced at the LHC \cite{Anchordoqui:2003ug,Meade:2007sz}. The main reasons are that not all of the energy of the partons is available for black hole formation and the parton distribution functions (PDFs)  tend to fall off very fast. 

The aim of this work is to study the production of quantum black holes (QBHs)  at the LHC. We define QBHs as the quantum analogs of ordinary black holes as their mass and Schwarzschild radius approach the quantum gravity scale. QBHs do not have semiclassical spacetimes and are not necessarily well-described by the usual Hawking temperature or black hole thermodynamics. In many respects they are perhaps more analogous to strongly coupled resonances or bound states than to large black holes. QBHs presumably decay only to a few particles, each with Compton wavelength of order the size of the QBH. It seems unlikely that they would decay to a much larger number of longer wavelength modes.

We shall assume that QBHs are defined by three quantities: their mass, spin and gauge charges. Importantly, QBHs can have a QCD, or color, charge. This is not in contradiction with confinement since the typical length scale of QCD, i.e., a Fermi, is much larger than the size of a QBH, e.g., TeV$^{-1}$. The formation and decay of a QBH takes place over a small spacetime region -- from the QCD perspective it is a short distance process, and hadronization occurs only subsequently. 

Our central assumptions are as follows.

\begin{itemize}

\item[I)] Processes involving QBHs conserve QCD and U(1) charges since local gauge symmetries are not violated by gravity. Note we make no similar assumption about global charges.
\item[II)] QBH coupling to long wavelength and highly off-shell perturbative modes is suppressed.
\end{itemize}
Assumption (II) is necessary so that precision measurements (e.g., of the anomalous magnetic moment of the muon \cite{Calmet:1976pu}) or, possibly, proton decay do not force the quantum gravity scale to be much larger than the TeV range. It is not implausible that a nonperturbative QBH state couples only weakly to long distance or highly off-shell modes, but strongly to modes of size and energy similar to that of the hole. This is analogous to results obtained for (B+L) violating processes in the standard model: (B+L) violation is exponentially small in low energy reactions, but of order one for energies above the sphaleron mass.

It is hard to imagine that (I) does not hold. Imagine a large Gaussian 3-sphere surrounding the spatial region where QBH formation and decay occurs. By causality, the total flux through this sphere is constant, implying conservation of charge. A consequence of assumption (I) is that QBHs can be classified according to representations of SU(3)$_c$ and U(1)$_{em}$. We will label QBH states as QBH$_c^q$. 

Note we did not list Lorentz invariance as one of the central assumptions. Our results will depend significantly on whether we require that QBH processes correspond to Lorentz invariant, local effective field theory operators (i.e., constructed from the usual standard model fields). We know of no argument in favor of this which is as robust as the one for conservation of gauge charges. The black hole production and decay take place over a small region of spacetime with Planckian volume. Whether or not this process can be matched to a local operator in an effective field theory description at larger length scales seems an open question. If quantum gravity does violate Lorentz invariance in QBH processes, we assume that assumption (II) above is sufficient to protect low energy physics (e.g., precision measurements) from contamination by these effects.

Finally, in order to obtain quantitative results we shall assume that QBH production cross-sections can be extrapolated from the cross-section obtained for semiclassical black holes \cite{Anchordoqui:2003ug} (see also \cite{Dimopoulos:2001hw,Giddings:2001bu} for earlier, less elaborated, cross-sections)
\begin{eqnarray}
\sigma^{pp}(s,x_{min},n,M_D) &=& \int_0^1 2z dz \int_{\frac{(x_{min} M_D)^2}{y(z)^2 s}}^1 du \int_u^1 \frac{dv}{v}  \\ \nonumber && \times F(n) \pi r_s^2(us,n,M_D) \sum_{i,j} f_i(v,Q) f_j(u/v,Q)
\end{eqnarray}
where $z=b/b_{max}$, $x_{min}=M_{BH,min}/M_D$,  $n$ is the number of extra-dimensions, $F(n)$ and $y(z)$ are the factors introduced by Eardley and Giddings (we use numerical values from \cite{Yoshino:2002tx}) and 
\begin{eqnarray}
r_s(us,n,M_D)=k(n)M_D^{-1}[\sqrt{us}/M_D]^{1/(1+n)}
\end{eqnarray}
where
\begin{eqnarray}
k(n) =  \left [2^n \sqrt{\pi}^{n-3} \frac{\Gamma(3+n)/2}{2+n} \right ]^{1/(1+n)},
\end{eqnarray}
and $M_D$ is the reduced Planck mass.
$M_{BH,min}$ is defined as the minimal value of black hole mass for which the semiclassical extrapolation can be trusted. Typically one expects that the construction of Eardley and Giddings holds for $M_{BH} \gg M_D$ and a semiclassical black hole will only form if, e.g., $M_{BH}\ge 3 M_D$.  For our numerical estimates  we use CTEQ5 PDFs for which an unofficial mathematica version is available on the webpage of the CTEQ collaboration and we take $Q\sim M_D$. We have fitted the functions $y(z)$ to the curves given in \cite{Yoshino:2002tx}. Note that there could be a suppression of the quantum cross-section relative to the extrapolated semiclassical one, i.e., the cross section is reduced dramatically as the black hole mass drops below $\sim 3 M_D$, but this would require the existence of some small dimensionless parameters characterizing strong gravitational scattering. We shall assume otherwise.

We do not expect that QBHs will have high angular momentum. The incoming partons are effectively objects which are extended in space-time, their typical size is fixed by $M_D^{-1}$ (i.e., due to a minimal length imposed by quantum gravity \cite{Calmet:2004mp}), which is also the interaction range of the semiclassical formation process in the limit of a quantum black hole. Thus, the impact parameter and hence the angular momentum of the QBH are small -- at impact parameter $M_D^{-1}$ the classical angular momentum would be order one at most. A classical black hole of this size with large angular momentum would have to spin at faster than the speed of light. Thus, the spin down process before the final explosion discussed in the context of semiclassical black holes does not take place here -- QBHs decay immediately to a small number of final states. 

Generically speaking, QBHs form representations of SU(3)$_c$ and carry a QED charge.
In our notation we denote the process of two partons $p_i$, $p_j$ forming a quantum black hole in the $c$ representation of SU(3)$_c$ and charge $q$ as: $p_i+p_j \to$ QBH$_c^q$. The following different transitions are possible at a proton collider:
\begin{itemize}
\item[a)] ${\bf 3} \times {\bf \overline 3}= {\bf 8} + {\bf 1}$ \\
\item[b)] ${\bf 3} \times {\bf 3}= {\bf 6} + {\bf \overline 3}$\\
\item[c)] ${\bf 3} \times {\bf 8}= {\bf 3} + {\bf \overline 6}+ {\bf 15}$\\
\item[d)] ${\bf 8} \times {\bf 8}= {\bf 1}_S + {\bf 8}_S+ {\bf 8}_A+{\bf 10} + {\bf \overline{10}}_A+ {\bf 27}_S$
\end{itemize}
Most of the time the black holes which are created carry a SU(3)$_c$ charge and come in different representations of SU(3)$_c$. This has important consequences for the production of QBHs. For example the production cross-section of a QBH$_1^0$ is given by
\begin{eqnarray}
\sigma^{pp}(s,x_{min},n,M_D) &=& \int_0^1 2z dz \int_{\frac{(x_{min} M_D)^2}{y(z)^2 s}}^1 du \int_u^1 \frac{dv}{v}  \\ \nonumber && \times F(n) \pi r_s^2(us,n,M_D)
\\ \nonumber &&
\left( \frac{1}{9} \sum_{i,j=q,{\bar q}} f_i(v,Q) f_{\bar j}(u/v,Q)
+\frac{1}{64}  f_g(v,Q) f_g(u/v,Q) \right)
\end{eqnarray}
where $i,j$ runs over all the quarks and anti-quarks subject to the constraint of QED charge neutrality,
and $f_q, f_g$ are the quark and gluon PDFs. For the production of a specific member (i.e., with specified color) of the octet QBH$_8^0$, one finds the same expression.


Since we know the total cross-section, at least semiclassically, it is straightforward to estimate the decay width in the same spirit of extrapolation. It is given by
\begin{eqnarray}
\Gamma({\rm QBH}_c^q \to p_1... p_f)\sim \left( 2 \pi  \left(\frac{1}{(2 \pi)^2}\right)^{(n_f-1)} \left(\frac{1}{2}\right)^{(n_f-1)} \right) \pi r_s^2 M_{BH}^3~~.
\end{eqnarray}
For quantum black holes one expects that the number of particles  $n_f$ in the final state is small, e.g., two or three. The three particle final state is strongly suppressed with respect to the two particle final state due to phase space. One thus typically has
\begin{eqnarray}
\Gamma = \frac{1}{4 \pi} \frac{M_{BH}^5}{M_D^4}
\label{width}
\end{eqnarray}
which for quantum black holes is of the order of $M_{BH}/4\pi \sim 80$ GeV for a quantum black hole with a mass of one TeV. Although consistent with our assumptions, the factor of $4 \pi$ in this width could be larger in reality; for example, we have neglected the sum over multiple decay channels. The actual decay width is model dependent. Another argument in favor of the decay of a quantum black hole to a two particle final state is that if the center of mass collision energy is lowered it should match the $2 \to 2$ cross-section with an exchange of a graviton. In studies of gravitational scattering by Amati, Ciafaloni and Veneziano \cite{Amati:2007ak}, evidence for an absorptive part of the forward amplitude is found near what would be the threshold of black hole formation using the Hoop Conjecture \cite{hoop}. This is consistent with our picture of quantum black holes as being gravitational bound states of, e.g., two particles.

A QCD-singlet quantum black hole which is also neutral under U(1)$_{em}$, denoted as QBH$_1^0$, will decay to any combination of higgs boson, leptons, quarks as well as gauge bosons and gravitons, e.g. QBH$_1^0 \to e^+ +e^-$, QBH$_1^0 \to e^+ + \mu^-$, QBH$_1^0 \to q_i +\bar q_i$ etc., as long as the global final state is neutral under QCD and U(1)$_{em}$. Because of the number of colored fermions in the standard model, most of the time the QBH$_1^0$ will decay to two jets. 

An octet black hole which is U(1)$_{em}$ neutral, QBH$^0_8$, can decay to a quark and an anti-quark of the opposite charge or to a gluon and a neutral particle such as a Z-boson or a photon. Further analysis of this channel depends on whether one imposes that there must exist a Lorentz invariant effective field theory description of black hole reactions. If we assume that Lorentz invariance is not violated, then transitions of the type $q_i +g \to$ QBH$_c^q \to q_k +q_j$, where $q_i$ are quarks and $g$ is a guon, will not take place as it is impossible to write down a Lorentz invariant local operator linking three fermions and a spin one gauge boson. 

A U(1)$_{em}$ charged triplet black hole QBH$_3^q$ can decay to a quark of charge $q$ and a gluon. These black holes will have two jets in the final state. Other decay modes of the QBH$_3^q$ which violate Lorentz conservation are for example quark+photon, quark+Z-boson, quark+graviton, quark+neutrino and quark+anti-neutrino. Similar considerations apply to black holes in higher representations:  QBH$_{10}^0$, QBH$_{\overline{10}}^0$,
QBH$_{27}^0$, QBH$_{\overline{6}}^q$, QBH$_{15}^q$,  QBH$_{6}^q$ and  QBH$_{\overline{3}}^q$.

Our discussion thus far has been fairly model independent, with the exception of the production cross-section for QBHs which in the case of the model of Randall and Sundrum receives a further suppression due to the warping of the extra-dimension \cite{Meade:2007sz}.  There are three types of models which can give rise to QBHs at the LHC. ADD and RS are well known and do not need to be reviewed. However there is a new four-dimensional model  \cite{Calmet:2008tn} which we shall refer to as CHR which we wish to describe in a few words. Even in four-dimensions, the Planck scale can be in the TeV range if there is a large hidden sector with some $10^{32}$ new particles which only interact gravitationally with the particles of the standard model. These new particles will lead to a renormalization of the Planck scale and gravity becomes strong at renormalization scale $\mu_\star$, where $M_P(\mu_\star)=\mu_\star$. For  $10^{32}$  new particles $\mu_\star \sim$ TeV.

We have calculated the inclusive production rate at the LHC of quantum black holes. The results are given in table 1.  As expected, quantum black hole processes dominate over semiclassical black holes. In ADD  and CHR we assumed that semiclassical black holes form already for $x_{min}=3$ whereas in RS we took $x_{min}=5$. In CHR, because of the large hidden sector, a non-negligible fraction of the quantum black holes decay invisibly. In ADD some missing energy will be emitted in the bulk via graviton decay of QBHs. In RS, because of the mass gap, most of the energy goes in the brane and QBHs thus decay visibly.

\begin{table}[tb]
\begin{center}
\begin{tabular}{|c|c|c|c|}
\hline
models & $\sigma$(p+p $\to$ any QBH)  in fb &$\sigma$(p+p $\to$ sc-BHs) in fb& 
  $\sigma$(p+p $\to$ m.e.) in fb \\
\hline
RS& $1.9 \times 10^6$& $ 151$ & $\sim$ none\\
\hline
ADD $n=5$& $9.5 \times 10^6$& $3.1 \times 10^4$& some\\
\hline
ADD $n=6$& $1.0 \times 10^7$& $3.2 \times 10^4$ &some\\
\hline
ADD $n=7$& $1.1 \times 10^7$& $2.9 \times 10^4$ & some\\
\hline
CHR  & $1 \times 10^5$  & $5 \times 10^3$ & $744$\\
\hline
\end{tabular}
\end{center}
\caption{\label{table1}Cross-sections for the production of quantum black holes and semiclassical (sc) black holes. The missing energy  (m.e.) component is also indicated. We take the reduced Planck mass to be 1 TeV and thus restrict our considerations to ADD with $n\ge 5$ since lower dimensional models with $M_D=1$ TeV are ruled out by astrophysical data. Note that the bound on the reduced Planck mass in four-dimensions is only of the order of 488 GeV \cite{Calmet:2008rv}.}
\end{table}

Due to conservation of color, most quantum black hole events at the LHC will give rise to two jets. However, the standard model background can be large. Two interesting signatures with less or no background are: proton+proton $\to$ QBH $\to$ lepton + anti-lepton of another generation and proton+proton $\to$ QBH $\to$ lepton + jet. The latter can only occur if we allow violation of baryon and lepton number (which is model dependent; for example these symmetries might be gauged). For example, consider p+p $\to$ QBH$^{-2/3}_{\bar{3}} \to l^- + \bar d$. If described by a local effective field theory, it could correspond to the operator ${\cal O} = \bar{u^c}_L d_L \bar{e}_L d_R$. The reaction with an anti-lepton in the final state can be mediated by $qqql$. Processes like p+p $\to$ QBH$^{1/3}_{\bar{3}} \to \gamma + \bar d$ would correspond to an operator connecting three fermions to a vector particle, which violates Lorentz invariance.

Since gravity is democratic (i.e., it couples equally to all flavors) we expect  that
$\sigma$(p+p $\to$ QBH $\to e$ + jet) $~=~ \sigma$(p+p $\to$ QBH $\to \mu$ + jet)$~=~\sigma$(p+p $\to$ QBH $\tau$+ jet), neglecting the masses of the fermions. We list some cross-sections with a lepton and a jet in the final state in table 2. The final state lepton can belong to the first, second or third generation. Some QBH$_{\bar 3}$ black holes will lead to a remarkable signature with a jet and a lepton back to back with high $p_T$. The lepton can be a neutrino, in which case the signature is missing energy with a high $p_T$ jet. Note that gauge bosons and the Higgs boson can appear in the final state: e.g., QBHs can decay to a Z and a jet or a Higgs boson + jet. The cross-section $\sigma$(p+p $\to$ Z+jet)  is equal to $3/2 \times \sigma$(p+p $\to \gamma$+jet). One can also have final states  involving missing energy, e.g., QBH $\to$ graviton+jet, $\nu$+jet and  $\bar \nu$+jet. Other interesting decay modes of QBHs involve a gluon and a photon in the final state. These can also appear in string theory as recently pointed out by Anchordoqui et al. \cite{Anchordoqui:2007da,Anchordoqui:2008hi}. Note that the width obtained in the calculation for decay of lowest massive Regge excitations of open strings \cite{Anchordoqui:2008hi} agrees with our result in eq. ({\ref{width}). As discussed in  \cite{Anchordoqui:2007da,Anchordoqui:2008hi}, the Lorentz conserving transitions $q+g \to$ QBH $\to \ q+ \gamma$ or 
$g+g \to$ QBH $\to \ g+ \gamma$ could lead to interesting signals, although the Standard Model background might be larger in these cases.

\begin{table}[tb]
\begin{center}
\begin{tabular}{|c|c|c|c|c|c|}
\hline
cross-sections in fb & CHR & RS& ADD $n=5$ & ADD $n=6$ & ADD $n=7$   \\
\hline
$\sigma$(p+p $\to$ QBH$^{4/3}_{\bar{3}} \to l^+ + \bar d )$ &  372&$5.8\times10^3$ &$3.3\times 10^4$ &$3.7\times 10^4$ &$4\times 10^4$ \\
\hline
$\sigma$(p+p $\to$ QBH$^{-2/3}_{\bar{3}} \to l^- + \bar d )$& 47& $734$&$3.7\times 10^3$ &$4\times 10^3$ & $4.2\times 10^3$
\\
\hline
 $\sigma$(p+p $\to$ QBH$^{1/3}_{\bar{3}} \to \nu_i + \bar d )$& 160& $2.5\times10^3$& $1.4\times 10^5$& $1.5\times 10^4$&$1.6\times 10^4$\\
 \hline
  $\sigma$(p+p $\to$ QBH$^{-2/3}_{\bar{3}} \to \nu_i + \bar u )$& 47&$734$ & $3.7\times 10^3$ &$4\times 10^3$ & $4.2\times 10^3$  \\
\hline
$\sigma$(p+p $\to$ QBH$^{-2/3}_{\bar{3}} \to \gamma+ \bar u )$& 47&$734$ & $3.7\times 10^3$ &$4\times 10^3$ & $4.2\times 10^3$  \\
\hline
$\sigma$(p+p $\to$ QBH$^{1/3}_{\bar{3}} \to \gamma + \bar d )$& 160& $2.5\times10^3$& $1.4\times 10^4$& $1.5\times 10^4$&$1.6\times 10^4$\\
 \hline
$\sigma$(p+p $\to$ QBH$^0_1$ $\to e^+ + \mu^-)$& 0& 93&$447$ &$491$& $511$ \\
\hline
\end{tabular}
\end{center}
\caption{\label{table2}Some possible final states in quantum black hole decay for the models CHR, RS and ADD. Gravity is democratic, one thus expects the same cross-sections for final states with any charged lepton combination. Note that if the neutrino is a Majorana particle the cross-section $\sigma$(p+p $\to$ QBH$^{-2/3}_{\bar{3}} \to \nu_i + \bar u )$ is $11/9$ times larger than what is given in the table, since one cannot differentiate $\nu$ from $\bar \nu$. If the neutrino is a Dirac type particle, then one has $\sigma$(p+p $\to$ QBH$^{-2/3}_{\bar{3}} \to \nu_i + \bar u )$ = $\sigma$(p+p $\to$ QBH$^{-2/3}_{\bar{3}} \to \bar \nu_i + \bar u )$. Note that we have summed over the polarization of the photon for the cross-sections $\sigma$(p+p $\to \gamma$+jet). The cross-section $\sigma$(p+p $\to$ Z+jet) = $3/2 \times \sigma$(p+p $\to \gamma$+jet).}
\end{table}

Another interesting signature of quantum black holes are decays of neutral, SU(3)$_c$ singlet holes to two leptons of different generations with opposite charge. These decays are highly suppressed in the CHR model since QBH$_1^0$ will decay invisibly in the hidden sector. However, in ADD and RS these signatures would be characteristic of quantum black holes. We have estimated the cross-sections $\sigma$(p+p $\to$ (neutral QBHs) $\to e^+ +\mu^-)$ and our results can be found in table 2. Again because of the universality of gravity we expect: $\sigma$(p+p $\to$ (neutral QBHs) $\to e^++\mu^-) ~=~ \sigma$(p+p $\to$ (neutral QBHs) $\to e^+ + \tau^-) ~=~ \sigma$(p+p $\to$ (neutral QBHs) $\to \tau^+ + \mu^-) ~=~ \sigma$(p+p $\to$ (neutral QBHs) $\to l^+ + l^-)$ where the l$^\pm$ can be any lepton.

\bigskip

{ \it Conclusions}

We have shown that although semiclassical black holes are unlikely to play a role at the LHC, quantum black holes which decay only to a few, most likely two, particles can lead to sizable cross-sections at the LHC. One striking signature is two leptons of different flavors and opposite electric charges in the final state. Another signature comes from the decay of quantum black holes which are in non trivial representations of SU(3)$_c$, resulting in a final state with a quark, leading to a jet, back to back with a lepton. Such events have little standard model background. 

From the theoretical point of view we have shown that very conservative assumptions about the dynamics of quantum black holes can help to make predictions for their decay modes. The calculated production cross-sections and decay rates of quantum black holes are such that they would generate numerous interesting events at the LHC.

\bigskip

\emph{Acknowledgments---}  We would like to thank Luis Anchordoqui, Haim Goldberg, Fabio Maltoni, Thomas Nunnemann and David Strom for helpful discussions. This work is supported in part by the
Department of Energy under DE-FG02-96ER40969 and  by the Belgian Federal Office for Scientific, Technical and Cultural Affairs through the Interuniversity Attraction Pole P6/11.

\bigskip

\emph{Source code}: Fortran source code, which can be used to compute QBH cross sections, and a detailed user guide, are available at 
http://duende.uoregon.edu/$\sim$hsu/qbh/.
See e.g. \cite{generators} for other black hole event generators. These
tend to focus on semiclassical black holes, although in some cases one
can specify a few (two) particle final state.

\bigskip
\bigskip



\bigskip


\end{document}